\documentclass[aps,prc,reprint,superscriptaddress,showpacs,amsmath,amssymb]{revtex4-2}
\usepackage{amsmath}
\usepackage{balance}
\usepackage{graphicx}
\usepackage{subfig}
\usepackage{dcolumn}% Align table columns on decimal point
\usepackage{bm}% bold math
\usepackage{caption}
\usepackage{hyperref}% add hypertext capabilities
\usepackage{multirow} 
\hypersetup{colorlinks=true, citecolor=blue, urlcolor=blue, linkcolor=blue}
\begin{document}
%\captionsetup[figure]{name={Fig.},labelsep=period,singlelinecheck=false} 
%\preprint{}

\title{ Non-extensive (3+1)-dimensional hydrodynamics for relativistic heavy-ion collisions}


\author{Jia-Hao Shi}
\affiliation{School of Physics and Information Technology, Shaanxi Normal University, Xi'an 710119, China}
\author{Zhi-Ying Qin}
\affiliation{School of Physics and Information Technology, Shaanxi Normal University, Xi'an 710119, China}
\author{Jin-Peng Zhang}
\affiliation{School of Physics and Information Technology, Shaanxi Normal University, Xi'an 710119, China}
\author{Jian Cao}
\affiliation{School of Physics and Information Technology, Shaanxi Normal University, Xi'an 710119, China}
\author{Ze-Fang Jiang}
\affiliation{Department of Physics and Electronic-Information Engineering, Hubei Engineering University, Xiaogan, Hubei 432000, China}
\affiliation{Institute of Particle Physics and Key Laboratory of Quark and Lepton Physics (MOE), Central China Normal University, Wuhan, Hubei 430079, China}
\author{Wen-Chao Zhang}
\email[]{wenchao.zhang@snnu.edu.cn}
%\thanks{}
\affiliation{School of Physics and Information Technology, Shaanxi Normal University, Xi'an 710119, China}
\author{Hua Zheng}
\affiliation{School of Physics and Information Technology, Shaanxi Normal University, Xi'an 710119, China}

\date{\today}

\begin{abstract}

A non-extensive (3+1)-dimensional hydrodynamic model for multi-particle production processes, NEX-CLVisc, is developed in the framework of CLVisc where the viscous corrections are turned off. It assumes that the non-extensive effects consistently exist in the initial conditions set by the optical Glauber model, the equation of state and the hadron kinetic freeze-out procedure. The model is then applied to simulate the pseudo-rapidity ($\eta$) distribution, the transverse momentum ($p_{\rm T}$) spectra and the $p_{\rm T}$-differential elliptic flow ($v_2$) of charged particles  in Pb-Pb collisions at $\sqrt{s_{\rm NN}}=$ 2.76 TeV and 5.02 TeV, respectively. It is found that the model can reasonably well reproduce the experimental data of the $\eta$ distribution and the charged-particle spectra in a $p_{\rm T}$ range up to 6–8 GeV/c. When compared with the ideal hydrodynamic model, the $p_{\rm T}$-differential $v_2$ of charged particles is suppressed in the NEX-CLVisc model, which is similar to that observed in the  hydrodynamic model with a shear viscous correction. Moreover, due to the lack of the viscous corrections and the event-by-event fluctuation, the model can only describe the $p_{\rm T}$-differential $v_2$ up to 3-4 GeV/c, which is smaller than its applicable range for the particle $p_{\rm T}$ spectra.

%Moreover, the applicable range for the model describing the $p_{\rm T}$-differential $v_2$ is up to 4 GeV/c, which is smaller than that indicated by the particle $p_{\rm T}$ spectra. 
 
\end{abstract}

%\keywords{}

\maketitle

\section{Introduction}\label{sect:intro}
Quantum-Chromodynamics (QCD) predicts that at high temperatures and energy densities there is a phase transition from the ordinary matter made of protons and neutrons to a hot and dense strongly interacting matter \cite{QGP1}. This matter is commonly denoted as quark-gluon plasma (QGP), where partons (quarks and gluons) are the dominant degrees of freedom \cite{QGP2, QGP3}. Such a phase transition would be expected in the early universe, at the center of compact stars, and in the initial stage of the ultra-relativistic heavy-ion collisions. The third possibility was firstly proposed by T.-D. Lee during the mid 1970s \cite{TDLee} and is currently being studied at the BNL Relativistic Heavy Ion Collider (RHIC) and CERN Large Hadron Collider (LHC).

Relativistic hydrodynamic model provides a simple picture of the space-time evolution of the QGP produced in high-energy nucleus-nucleus (AA) collisions \cite{hydro_0,hydro_1}. It gives a reasonable description of the experimental data on various observables such as the pseudo-rapidity ($\eta$) distribution, the transverse momentum ($p_{\rm T}$) spectra, and the $p_{\rm T}$-differential elliptic flow ($v_2$) of charged hadrons in the low $p_{\rm T}$ region \cite{hydro1, hydro2, hydro3, hydro4, hydro5, hydro6, hydro7}. The model-to-data comparison shows that QGP in AA collisions exhibits properties of nearly-perfect fluid with a small shear viscosity per entropy density \cite{shear_1, shear_2, shear_3}. 

For the  hydrodynamic models, a local thermal equilibrium is assumed and the Boltzmann-Gibbs (BG) statistics is applied. With such an assumption, the typical validity range of the hydrodynamic description on the $p_{\rm T}$ spectra of charged hadrons is around 2-3 GeV/c. However, for a multiparticle production
process in relativistic AA collisions, conditions that lead to BG statistics are satisfied only approximately at best, since the hadronizing systems would experience strong intrinsic fluctuations and long-range correlations \cite{temp_fluct,temp_fluct_1,non_ext_hydro,q_cent_dependence_1}. Thus, instead of local equilibrium state, some kind of stationary state near the equilibrium ($q$-equilibrium) is expected to be formed \cite{stationary_state_1, stationary_state_2}. In such a circumstance, some quantities such as the $p_{\rm T}$ spectra will develop power-law tailed rather than exponential distributions in the high $p_{\rm T}$ region. This phenomenon can be incorporated in terms of a non-extensive statistics without going into deeper dynamical considerations about the source of the fluctuations.

The non-extensive statistics was first proposed by Tsallis \cite{nex1}. It replaces the usual exponential function in the BG statistics with the $q$-exponential function,
 \begin{equation}\label{eq:Tsallis_dist}
 	\textrm{exp}_q(x)=\begin{cases}
		[1+(1-q)x]^{(1/(1-q))}, x\leq0\\
		[1+(q-1)x]^{(1/(q-1))}, x>0
             \end{cases},
 \end{equation}
where  $ q $ is the non-extensive parameter representing the extent of deviation from thermal equilibrium. If $q$ approaches to $ 1 $, the Tsallis statistics degenerates to the BG statistics. When taking into account the quantum effects of particles, the Tsallis distribution is generalized to the following form  \cite{nex17, nex18},
\begin{equation}\label{eq:non_extensive_quantum}
	f(x)=\frac{\textrm{exp}_q(x)}{1\pm\textrm{exp}_{q}(x)},
\end{equation}
where the plus and minus signs, respectively, refer to fermions and bosons.

In recent years, the Tsallis statistics has been widely applied to describe the multi-particle production processes in high energy proton-proton \cite{nex17, nex18, Tsallis_10, Tsallis_11, Tsallis_12, Tsallis_13, Tsallis_14, Tsallis_15, Tsallis_16} and AA \cite{Tsallis_1, Tsallis_2, Tsallis_3, Tsallis_4, Tsallis_5, Tsallis_6, Tsallis_7, Tsallis_8, Tsallis_9} collisions. In these applications, it is observed that $q$ is larger than 1. The deviation of $q$ from unity represents the intrinsic fluctuations of the temperature \cite{temp_fluct} or of the mean value of the charged-particle multiplicity \cite{mult_fluct} in the hadronizing system. However, in most of these applications the Tsallis distribution or the Tsallis-extended blast-wave model is simply fitted to the particle $p_{\rm T}$ spectra, which could get a biased  result in the case of high-energy AA collisions where the fluid dynamics plays an important role \cite{eos}. As described in Ref. \cite{q_hydro_1}, 
hydrodynamic behavior does not require full thermalization. Isotropization of parton momenta in local fluid rest frames is enough. Thus, far-from-equilibrium ($q$-equilibrium) evolution can still obey the hydrodynamic equations of motion, provided that the system is locally isotropized in the co-moving frame of the fluid \cite{q_hydro_2}. Due to this reason, in Ref. \cite{non_ext_hydro}, the authors proposed a (1+1)-dimensional non-extensive hydrodynamic ($q$-hydrodynamic) model where the $q$-corrections were applied to the initial conditions, the equation of state, and the Cooper-Frye prescription of the kinetic freeze-out \cite{Cooper_freeze}. In Ref. \cite{eos}, the authors developed a (2+1)-dimensional $q$-hydrodynamic model with the application of the $q$-corrections to the equation of state and the kinetic freeze-out procedure. However, the former model ignores the transverse expansion of the fluid, so it is impossible to study the anisotropic flow of identified particles. The latter model assumes a boost invariant in the longitudinal expansion, so it is unable to investigate the single-particle one-dimensional distributions in longitudinal phase-space.

In this work,  a non-extensive (3+1)-dimensional hydrodynamic model, NEX-CLVisc, is developed on the basis of the CLVisc framework \cite{clvisc}. In order to elucidate the pure effect of non-extensive statistics, the viscous corrections have been turned off. In the model, the $q$-corrections are consistently taken into account in the initial conditions, the equation of state, and the kinetic freeze-out procedure. The intrinsic fluctuations in different stages of the hydrodynamic evolution are expected to be different, thus the parameters $q$ for the initial conditions, the equation of state, and the kinetic freeze-out procedure could be different. However, for simplicity, we restrict ourselves to use the same $q$ values for all stages of the collision. We then use this model to calculate the $\eta$ distribution, the $p_{\rm T}$ spectra and the $p_{\rm T}$-differential $v_2$ of charged particles and confront our results with the experimental data in Pb-Pb collisions at $\sqrt{s_{\rm NN}}=$ 2.76 TeV and 5.02 TeV, respectively. Moreover, we have compared our results with those obtained from the ideal hydrodynamic model with the BG statistics.

The organization of this paper is as follows. In section \ref{sect:initial_condition}, we propose a non-extensive version of the initial conditions. In section \ref{sect:eos}, we construct a QCD equation of state with the non-extensive statistics. In section \ref{sect:freeze_out}, the $q$-correction is applied to the  freeze-out procedure and the calculated charged-particle $\eta$ distributions, $p_{\rm T}$ spectra and elliptic flow are presented. The discussions and conclusions are given in section \ref{sec:conclusion}.

\section{Non-extensive initial conditions}\label{sect:initial_condition}
In CLVisc, there are several models to set the initial conditions for the hydrodynamic evolution.  These are the optical Glauber model \cite{optical_Glauber}, the Trento model \cite{Trento} and the a-multiphase-transport (AMPT) model \cite{AMPT}. In this work, the optical Glauber model is adopted.

In this model, with the parameterization of the nucleon density as the Woods-Saxon distribution, the nucleus thickness function $T(x,y)$ is written as
\begin{equation}\label{thickness_funct}
	T(x,y)=\int_{-\infty}^{\infty}\mathrm{d}z\frac{\rho_0}{1+e^{(\sqrt{x^2+y^2+z^2}-R)/d}},
\end{equation}
where $\rho_0$ is the nucleon density in the center of the nucleus, $R$ is the nuclear radius, $d$ is the skin depth, $ x, y, z $ are the space coordinates of nucleons. The Woods-Saxon parameters of Pb used in this work are listed in Table \ref{tab:Pb_parameters}.

\begin{table}[h]
\caption{Parameters of the Woods-Saxon distribution for the Pb nucleus \cite{mc}.}\label{tab:Pb_parameters}
	\begin{ruledtabular}
	\begin{tabular}{ccccc}
		Nucleus & $ A $ & $ n_0\ [1/\mathrm{fm}^3] $ & $R$ [fm] & $d$ [fm]\\
		\colrule
		Pb & 208 & 0.16 & 6.62 & 0.546\\
	\end{tabular}
    \end{ruledtabular}
\end{table}

The thickness functions for the two nuclei propagating along the $\pm\hat{z} $ direction with an impact parameter $\bm{b}$ are,
\begin{equation}\label{4}
	T_+(\bm{s})=T(\bm{s}+\bm{b}/2),\quad T_-(\bm{s})=T(\bm{s}-\bm{b}/2),
\end{equation}
where $\bm{s}$=$(x,y)$ is the coordinate in the transverse plane,  $\bm{s}\pm\bm{b}/2 $ represent the displacements of the flux tube with respect to the centers of the target and the projectile nuclei in the transverse plane, respectively. The density of wounded nucleons in the transverse plane is then given by \cite{wounded_nucleons}
\begin{equation}\label{eq:wn}
\begin{split}
    n_{\rm WN}(x,y)&=T_+(x,y)\left\{1-\left[1-\frac{\sigma_{\rm NN}T_-(x,y)}{A}\right]^{A}\right\}\\
     &+T_-(x,y)\left\{1-\left[1-\frac{\sigma_{\rm NN}T_+(x,y)}{A}\right]^{A}\right\},
     \end{split}
\end{equation}
where $ A $ is the mass number of the nuclei and $ \sigma_{\rm NN} $ is the inelastic nucleon-nucleon cross section. $ \sigma_{\rm NN} $ is taken as $ 61.8\ \mathrm{mb} $ at $ \sqrt{s_{\rm NN}}=2.76\ \mathrm{TeV} $ and $ 67.6\ \mathrm{mb} $ at $ \sqrt{s_{\rm NN}}=5.02\ \mathrm{TeV} $ \cite{mc}.
The number of binary nucleon–nucleon collisions in the transverse plane is 
\begin{equation}\label{eq:bn}
	n_{\rm BC}(x,y)=\sigma_{\rm NN}T_+(x,y)T_-(x,y).
\end{equation}

It is assumed that a large fraction of the initial energy deposition is due to the soft processes that is proportional to the number of wounded nucleons, and a small fraction comes from the hard processes that is  proportional to the number of binary collisions \cite{wounded_nucleons_1}. Thus the initial energy density profile in the transverse direction is given by  \cite{energy_density,energy_density_1,energy_density_2}
%\begin{equation}\label{eq:energy_density}
	%\varepsilon(x,y,\eta_s)=KW(x,y)H(\eta_s).
%\end{equation}
\begin{equation}\label{8}
	W(x,y)=\frac{(1-\alpha)n_{\rm WN}(x,y)/2+\alpha n_{\rm BC}(x,y)}{(1-\alpha)n_{\rm WN}(0,0)/2+\alpha n_{\rm BC}(0,0)|_{\bm{b}=0}},
\end{equation}
where the collision hardness parameter $ \alpha$ is assumed to be  energy independent and is set to be  0.05 \cite{hardness_parameter}. In CLVisc, in order to describe the plateau structure of the rapidity distributions of emitted hadrons, the energy density profile in the longitudinal direction is parameterized as \cite{energy_density,energy_density_1,energy_density_2}
\begin{equation}\label{eq:energy_density_longitudinal}
	H(\eta_s)=\exp\left[-\frac{(|\eta_s|-\eta_w)^2}{2\sigma^2_{\eta}}\theta(|\eta_s|-\eta_w)\right],
\end{equation}
where $\eta_s$ is the longitudinal space-time rapidity, $ \theta $ is a step function, $ \eta_w $ determines the extension of the central rapidity plateau and $ \sigma_{\eta} $ is the width of the Gaussian fall-off at large rapidity.

It is expected that there exist some intrinsic fluctuations of the initial energy densities in the transverse plane and along the longitudinal direction. Since we use the smooth initial condition from the optical Glauber model, the event-by-event fluctuations in the transverse plane are ignored in this work. As done in the (1+1)-dimensional  $q$-hydrodynamic model \cite{non_ext_hydro}, we only consider the fluctuations in the longitudinal direction and modify the longitudinal energy density profile in Eq. (\ref{eq:energy_density_longitudinal}) as 
\begin{equation}\label{eq:q_energy_density_longitudinal}
	H_q(\eta_s)=\exp_q\left[-\frac{(|\eta_s|-\eta_w)^2}{2\sigma^2_{\eta}}\theta(|\eta_s|-\eta_w)\right].
\end{equation}
With the combination of the energy density profiles in the transverse and longitudinal directions, the non-extensive initial energy density at the hydrodynamic starting proper time $\tau_0$ is written as 
\begin{equation}\label{eq:initial_condition}
	\varepsilon_{q}(x,y,\eta_s)=\varepsilon_{0} W(x,y)H_{q}(\eta_s),
\end{equation}
where $ \varepsilon_{0} $ is the maximum energy density. For both extensive and non-extensive hydrodynamic evolution in Pb-Pb collisions at $\sqrt{s_{\rm NN}}=$ 2.76 TeV and 5.02 TeV, $\tau_0$ is set to be 0.6 fm/c.
%determined by the particle yield in the most central AA collisions. 

The impact parameters for different centrality bins in Pb-Pb collisions are given by the following formula \cite{impact_parameter},
\begin{equation}\label{Eq:impact_parameter}
	b = \sqrt{c} \times b_{\rm max},
\end{equation}
where $ c $ refers to the centrality percentile and $ b_{\rm max} $ is assumed to be
\begin{equation}\label{30}
	b_{\rm max} = R_A + R_B + f \times d,
\end{equation}
with $R_A$ ($R_B$) being the radius of the nuclear $A\ (B)$, $d$ describing the tail of the nuclear density profile. In this work, $ R_A=R_B=6.62$ fm, $ d = 0.546$ fm and $ f = 6.15 $. With the number of wounded nucleons $N_{\rm WN}(b)$ as the weight function of the impact parameters, the average impact parameter in one given centrality bin is evaluated as \cite{clvisc}
\begin{equation}\label{eq:average_IP}
	\langle b \rangle=\frac{\int_{b_{\rm low}}^{b_{\rm high}} b^2 N_{\rm WN}(b)\mathrm{d}b}{\int_{b_{\rm low}}^{b_{\rm high}} b N_{\rm WN}(b)\mathrm{d}b},
\end{equation}
where $b_{\rm low}$ ($b_{\rm high}$) is the lower (upper) limit of the impact parameter in that centrality bin. These parameters are shown in Table \ref{tab:impact_para}.

\begin{table}[h]
\caption{Average impact parameters for different centralities in Pb-Pb collisions  at both $ \sqrt{s_{\rm NN}}$= 2.76 TeV and 5.02 TeV.}\label{tab:impact_para}
	\begin{ruledtabular}
	\begin{tabular}{ccccc}
		  &  0-5$\% $ & 5-10$\%$ & 10-20$\%$ & 20-30$\%$\\
		\colrule
		$b$ (fm) & 2.43 & 4.50 & 6.33 & 8.22\\
	\end{tabular}
    \end{ruledtabular}
\end{table}

\section{$q$-modified Equation of state} \label{sect:eos}
For an ideal fluid with zero baryon density, the hydrodynamic equation of motion is given by the conservation law,
\begin{equation}\label{eq:energy_momentum_cons}
	\partial_{\mu}T^{\mu\nu}=0,
\end{equation}
where $T^{\mu\nu}$ is the extensive energy-momentum tensor. In the non-extensive statistics, the $q$ version of the local energy-momentum conservation is written as
\begin{equation}\label{eq:energy_momentum_cons}
	\partial_{\mu}T^{\mu\nu}_q=0,
\end{equation}
where $T^{\mu\nu}_q$ is the non-extensive energy-momentum tensor and will be explained and detailed in the following subsection. The conservation law in Eq. (\ref{eq:energy_momentum_cons}) contains four independent equations. However, there are five thermodynamical variables, i.e. the $q$-modified energy density $ \varepsilon_q $ and pressure $ P_q $,  and the three components of the flow vector, $ v_x $, $ v_y $, and $ v_z $. For a given initial condition, in order to determine the space-time evolution of these variables, it is necessary to construct an equation of state (EOS) relating $ \varepsilon_q $ and  $ P_q $.
\subsection{The Model}

In the  kinetic theory, $T^{\mu\nu}_q$ of the system in the hadronic or the QGP phase can be written as \cite{non_ext_hydro, eos}
\begin{equation}\label{eq:T_mu_nu}
	T_q^{\mu\nu}=\sum_i\frac{g_i}{(2\pi)^3}\int\frac{\mathrm{d}^3p}{E_i}p_i^{\mu}p_i^{\nu}f^q_i(E_i/T),
\end{equation}
where  $g_i$, $E_i$ and $p_i$ are, respectively,  the degeneracy factor, the energy and the four momenta of the  particle type $i$, $T$ is the temperature of the system, $f_i(E_i/T)$ is the non-extensive phase-space distribution function with the consideration of the quantum statistics (see Eq. (\ref{eq:non_extensive_quantum})). The decomposition of the $q$-modified $T_q^{\mu\nu}$ is similar as that of the $T^{\mu\nu}$ in the extensive case and  can be done in terms of the $q$-modified energy density and pressure, $\varepsilon_q$ and $P_q$, by using the hydrodynamic flow $u^{\mu}$ \cite{non_ext_hydro,eos}, 
\begin{equation}\label{eq:T_mu_nu_1}
	T_q^{\mu\nu}=(\varepsilon_q+P_q)u^{\mu}u^{\nu}-P_qg^{\mu\nu}, 
\end{equation}
where 
\begin{align}\label{eq:energy_density}
	\varepsilon_q&=T_q^{\mu\nu}u_{\mu}u_{\nu}=\sum_i\frac{g_i}{(2\pi)^3}\int\mathrm{d}^3pE_if_i^q,\\
	P_q&=-\frac{1}{3}T_q^{\mu\nu}\Delta_{\mu\nu}=\sum_i\frac{g_i}{(2\pi)^3}\int\frac{\mathrm{d}^3p}{3E_i}\bm{p}_i^2f_i^q,\label{eq:pressure}
\end{align}
with $\Delta_{\mu\nu}=g^{\mu\nu}-u^{\mu}u^{\nu}$ being the projection operator. It is found in the local rest frame of the fluid  $T_q^{\mu\nu}$ has the form $T_q^{\mu\nu}=\textrm{diag}(\varepsilon_q,P_q,P_q,P_q)$, which is identical to the ideal fluid. Moreover,  for the baryon-free case, the  thermodynamic relations still hold in the non-extensive statistics \cite{non_ext_hydro},
\begin{equation}\label{eq:thermo_relation}
	T s_q=\varepsilon_q+P_q,
\end{equation}
and
\begin{equation}\label{eq:thermo_relation_1}
	\frac{\partial P_q}{\partial T}=s_q,
\end{equation}
where $s_q$ is the $q$-modified  entropy density \cite{nex17}, 
\begin{equation}\label{eq:entropy_density}
		s_q=\sum_i\frac{g_i}{(2\pi)^3}\int\mathrm{d}^3p[f_i^q\textrm{ln}_q f_i^q\pm(1\mp f_i)^q\textrm{ln}_q(1\mp f_i)],
\end{equation}
with the upper and lower signs, respectively, referring to the fermions and bosons.
%With the $q$ version of thermodynamic relations in Eq. (\ref{eq:thermo_relation}), Eq. (\ref{eq:energy_momentum_cons}) implies that the $q$-entropy current $s^{\mu}_q(x)=s_q(x)u^{\mu}(x)$ is conserved, i.e. $\partial_{\mu}s^{\mu}_q=0$.
%\begin{equation}\label{eq:entropy_cons}
	%\partial_{\mu}s^{\mu}_q=0,
%\end{equation}
%where $s^{\mu}_q(x)=s_q(x)u^{\mu}(x)$.

The equations of state in the hadronic and QGP phases are then, respectively, constructed by embedding the $q$-correction to the hadron resonance gas (HRG) model \cite{HRG} and parton resonance gas model \cite{eos}. At low temperatures, the HRG model describes the hadronic EOS quite successfully \cite{HRG_1}.  In order to match the pressure at the hadronic phase ($P^{\rm had}_q(T)$) with that at the QGP phase ($P^{\rm QGP}_q(T)$), the following procedure is adopted \cite{eos}: 
\begin{equation}\label{eq:mix_phase_pressure_1}
P_q(T)=P^{\rm had}_q(T),
\end{equation}
for $T\leq T_f$ and 
\begin{equation}\label{eq:mix_phase_pressure_2}
P_q(T)=S(T)P^{\rm had}_q(T_f)+(1-S(T))P^{\rm QGP}_q(T),
\end{equation}
for $T> T_f$ with  $S(T)=\textrm{exp}(-c(T-T_f))$ being an exponential damping function which ensures the energy-momentum conservation at the kinetic freeze-out \cite{eos}. In $S(T)$, the coefficient $ c $ is chosen to satisfy the constraint
\begin{equation}\label{22}
	\frac{\partial P_q}{\partial T}(T_f)=\frac{\partial P_q^{\rm had}}{\partial T}(T_f),
\end{equation}
which ensures a smooth and continuous matching of the equation of state in the mixed phase with that of the hadronic phase at the connection point defined with the kinetic freeze-out temperature $T_f$.

\subsection{Numerical results}
\begin{figure}[h]
	\centering
	\includegraphics[scale=0.4]{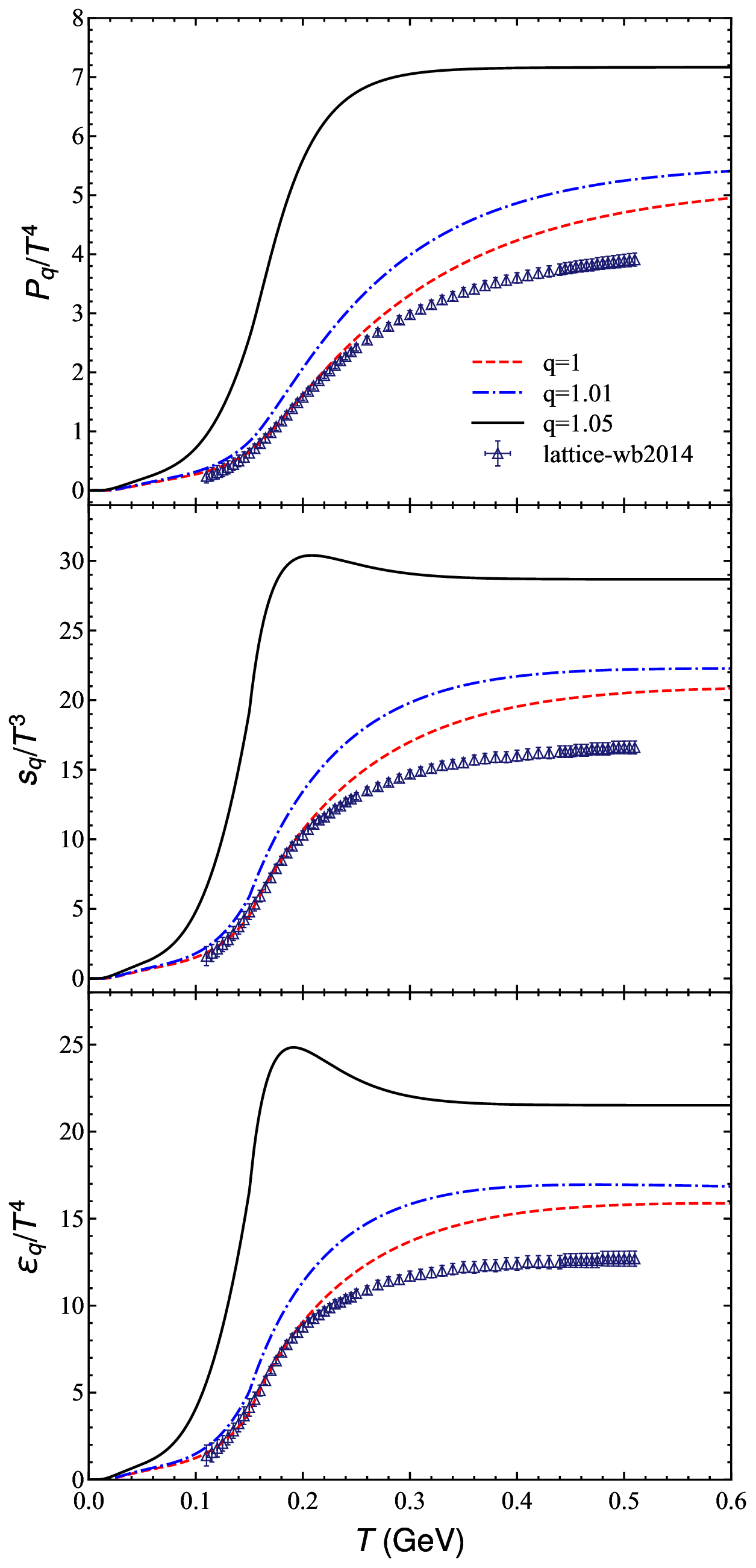}
	%\captionsetup{justification=raggedright}
	\caption{\label{fig:eos} $ P_q/T^4 $, $ s_q/T^3 $ and $ \varepsilon_q/T^4 $  as a function of $T$ at $ q=1 $ (solid), $ 1.01 $ (dashed), and $ 1.05 $ (dash-dotted) with $ T_f= 150$ MeV. Also presented are the results from the LQCD \cite{LQCD}.}
\end{figure}

In the simulation, the QGP phase consists of $u$, $d$, $s$ quarks and gluons. In the hadronic phase, hadrons and hadron resonances composed of $u$, $d$ and $s$ quarks with masses below 2 GeV/c$^2$ from the Particle Data Group \cite{pdg} are considered. Figure \ref{fig:eos}
displays the  dimensionless $ P_q/T^4 $,  $ s_q/T^3 $ and $ \varepsilon_q/T^4 $ as a function of temperature with different values of $q=$ 1, 1.01, and 1.05. For illustration, the kinetic freeze-out temperature is set to be 150 MeV. Also presented in the figure are the results from the 2+1 flavor Lattice QCD (LQCD) \cite{LQCD}. Their qualitative behaviors can be summarized as follows.

(i) In both the hadronic and QGP phases, the pressure, the energy density and the entropy density increase with the increase of $q$. This can be explained as follows. At fixed temperature, the particle yield in the Tsallis distribution is always larger than that in the Boltzmann one if $q > 1$. 

(ii) The pressure in the hadronic phase increases more rapidly with $q$ than that in the QGP phase, indicating that a large $q$ may be disfavored by
the thermodynamic conditions $\partial{P_q}/\partial{T}>$ 0 and $\partial{\varepsilon_q}/\partial{T}>$ 0 around $T_f$.

(iii) When $q=1.05$, $s_q/T^3 $ has an obvious peak just above $T_f$. The existence
of this peak is a consequence of the rapid increase of $P_q(T)$ around the crossover region. A similar peak is observed in the dependence of $ \varepsilon_q/T^4 $ on temperature.

It is found that our result with $q=1$ gives the best agreement with the LQCD data up to $T\sim 200$ MeV. At high temperature, as the partons are deemed as an ideal parton gas, our result approaches the Stefan-Boltzmann limit. However, in LQCD, partons interact with each other, which results deviations from the limit \cite{QGP2}. 

\section{KINETIC FREEZE-OUT UNDER NON-EXTENSIVE STATISTICS} \label{sect:freeze_out}
After the hydrodynamic evolution, the QGP will be converted into observed secondaries. A $q$-extended version of the Cooper-Frye formula \cite{Cooper_freeze} is employed at a freeze-out hyperspace $\Sigma$. The invariant momentum spectrum of hadrons reads 
\begin{equation}\label{eq:freeze_out_spectra}
	E\frac{\mathrm{d^3}N_i}{\mathrm{d}^3p}=\frac{g_i}{(2\pi)^3}\int_{\Sigma}f^q_{i} p^{\mu}d\sigma_{\mu}.
\end{equation}
where $p^{\mu}$ is the four-momentum of the emitted hadron, $d\sigma_{\mu}$ is the normal vector to the freeze-out hypersurface. As shown in Table \ref{tab:freeze_out_time}, with the increase of the non-extensive parameter $ q $, both  the lifetime of the fireball and the size of freeze-out hypersurface decreases. This could be understood as follows. From the bottom panel of Fig. \ref{fig:eos}, we observe that with a larger $q$ the initial temperature would be smaller for a given energy density, which will lead to a shorter lifetime and a smaller size of the fireball.  We consider the contribution of resonance decay according to Ref. \cite{resonance} and the list of decay channels is taken from  Ref. \cite{particle_list}.

\begin{table}[h]
\caption{The kinetic freeze-out time (in units of fm/c) for $q=1$ and $q=1.05$ at different centralities in Pb-Pb collisions  at $ \sqrt{s_{\rm NN}}$= 2.76 TeV and 5.02 TeV, respectively. } \label{tab:freeze_out_time}
\begin{ruledtabular}
\begin{tabular}{ccccc}
\multirow{2}{*}{} & \multicolumn{2}{c}{2.76 TeV } & \multicolumn{2}{c}{5.02 TeV} \\
   &  $ \tau(q=1) $  & $ \tau(q=1.05) $ & $ \tau(q=1) $  & $ \tau(q=1.05) $         \\ 
\colrule
0-5$\%$               & 14.32         & 9.23        & 14.71    & 9.55            \\
5-10$\%$          & 13.26      & 8.52      & 13.64 & 8.81            \\
10-20$\%$                         & 12.00        & 7.68       & 12.37   & 7.95            \\
20-30$\%$                         & 10.43        & 6.65        & 10.78    & 6.91               
\end{tabular}
\end{ruledtabular}
\end{table}

\subsection{Pseudo-rapidity density distributions}

The values of $\varepsilon_{0}$, $ \alpha$, $\eta_w $ and $ \sigma_{\eta} $ in the initial conditions for the  non-extensive and extensive hydrodynamic evolution in Pb-Pb collisions at $\sqrt{s_{\rm NN}}$ = 2.76 TeV and 5.02 TeV are listed in Table \ref{Pb_parameters2}, respectively. They are determined by the fit of the model to  the experimental  charged-particle pseudo-rapidity density ($d N_{\rm ch}/d\eta$) distribution in the middle rapidity region at the most central collisions. In this table, at a given energy, the parameters for the non-extensive case are the same as those for the extensive case, except for $\varepsilon_{0}$ whose value  in the former case is larger than that in the latter case. As described in Ref. \cite{non_ext_and_ext_comp}, at a fixed temperature, the non-extensive distribution is always larger than the extensive one if $q>1$. However, since the size of the freeze-out hypersurface for the former case is smaller than the latter case,  in order to keep the particle yields the same, the non-extensive statistics  leads to larger values of $\varepsilon_{0}$. At other centrality bins, the initial conditions are determined by varying the impact parameter in Eq. (\ref{eq:initial_condition}) according to Table \ref{tab:impact_para}.

\begin{table}[h]
\caption{Parameters of initial energy densities for Pb-Pb collisions  at $ \sqrt{s_{\rm NN}}$= 2.76 TeV and 5.02 TeV, respectively. }\label{Pb_parameters2}
\begin{ruledtabular}
\begin{tabular}{ccccc}
\multirow{2}{*}{} & \multicolumn{2}{c}{2.76 TeV} & \multicolumn{2}{c}{5.02 TeV} \\
           & $ q=1 $  & $ q=1.05 $ & $ q=1 $  & $ q=1.05 $         \\ 
\colrule
%$\tau_0(\mathrm{fm}/c)$                 & 0.6         & 0.6        & 0.6    & 0.6            \\
$ \varepsilon_{0}(\mathrm{GeV/fm^3}) $  & 111.8      & 133.24      & 138.5 & 165.19            \\
$ \alpha $                              & 0.05        & 0.05       & 0.05   & 0.05            \\
$ \eta_w $                              & 2.0         & 2.0        & 2.2    & 2.2            \\
$ \sigma_{\eta} $                       & 1.8         & 1.8        & 1.8    & 1.8           
\end{tabular}
\end{ruledtabular}
\end{table}

\begin{figure}[htbp]
	\centering
	
	%\captionsetup{justification=raggedright}
	\includegraphics[scale=0.39]{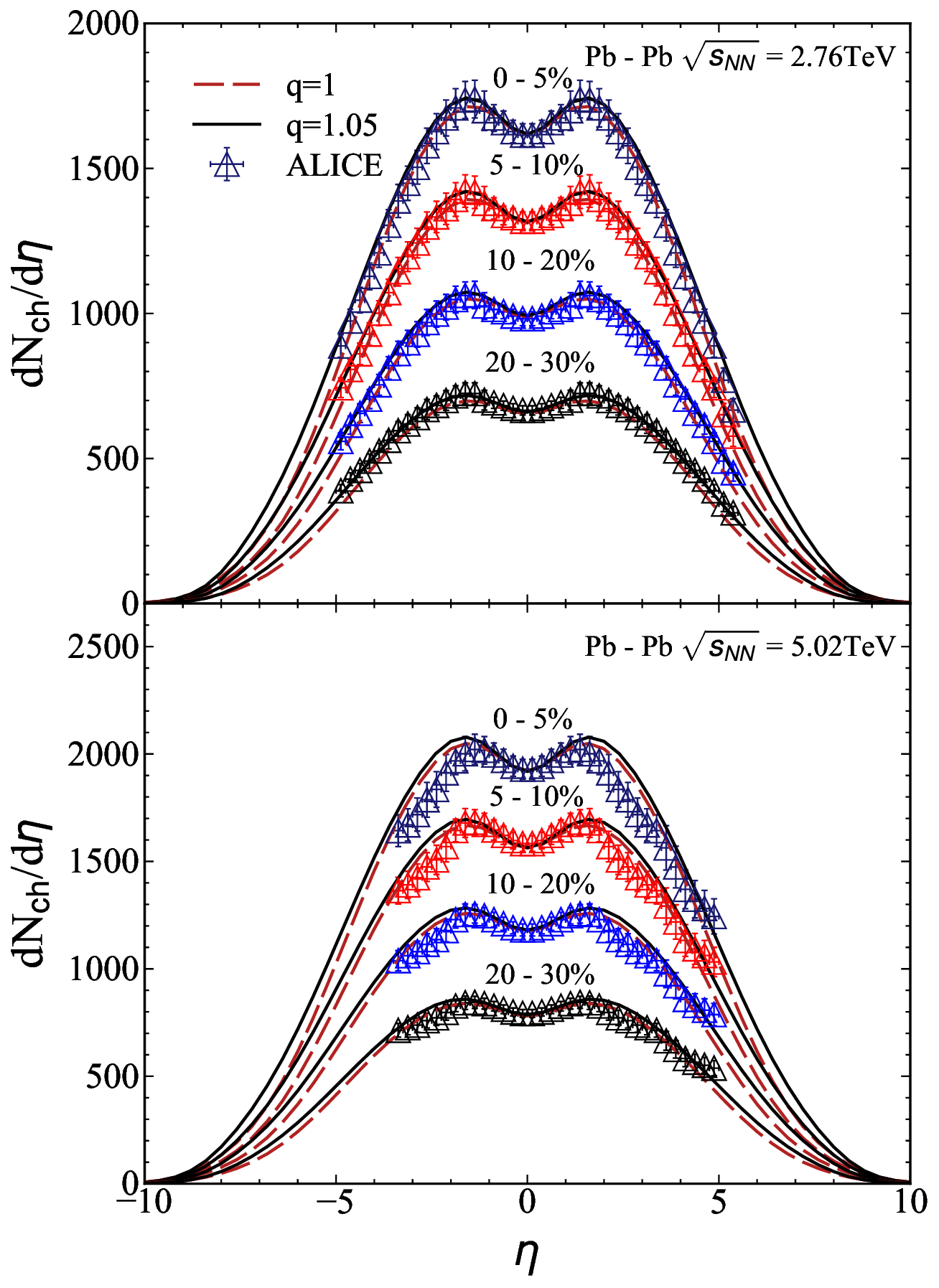}
	\caption{Upper (Lower) panel: pseudo-rapidity density distributions for charged particles at different centralities in Pb-Pb collisions at $ \sqrt{s_{\rm NN}} $ = 2.76 (5.02) TeV. The solid (dashed) curves are the results from the NEX-CLVisc model with $q=1.05$ ($q=1$). The triangle-up symbols are the experimental data taken from Refs. \cite{lhc_1, lhc_2}.}\label{fig:pseudo_dist}
\end{figure}

Figure \ref{fig:pseudo_dist} presents the pseudo-rapidity density distributions of charged particles from the NEX-CLVisc model with $q=1.05$ (solid curves)  and $q=1$ (dashed curves) at 0-5\%, 5-10\%, 10-20\%, and 20-30\% centralities in  Pb-Pb collisions at $ \sqrt{s_{\rm NN}} $ = 2.76 TeV and 5.02 TeV, comparing with experimental data (triangles-up). It is observed that the $d N_{\rm ch}/d\eta$ distributions are sensitive to $q$, which is entirely due to the $q$-dependence of the initial energy density in Eq. (\ref{eq:q_energy_density_longitudinal}). In the mid-rapidity region, both the extensive and non-extensive simulations can well describe the experimental data. For Pb-Pb collisions at $ \sqrt{s_{\rm NN}} $ = 2.76 TeV, the extensive simulation underestimates the data at large rapidities while the non-extensive simulation gives a reasonable description of the data. This trend becomes more pronounced for mid-peripheral collisions. For Pb-Pb collisions at $ \sqrt{s_{\rm NN}} $ = 5.02 TeV, both the extensive and non-extensive simulations overestimate the data in the large rapidity region.

\subsection{Transverse momentum spectra}
\begin{figure}[htbp]
	\centering
	
	%\captionsetup{justification=raggedright}
	\includegraphics[scale=0.39]{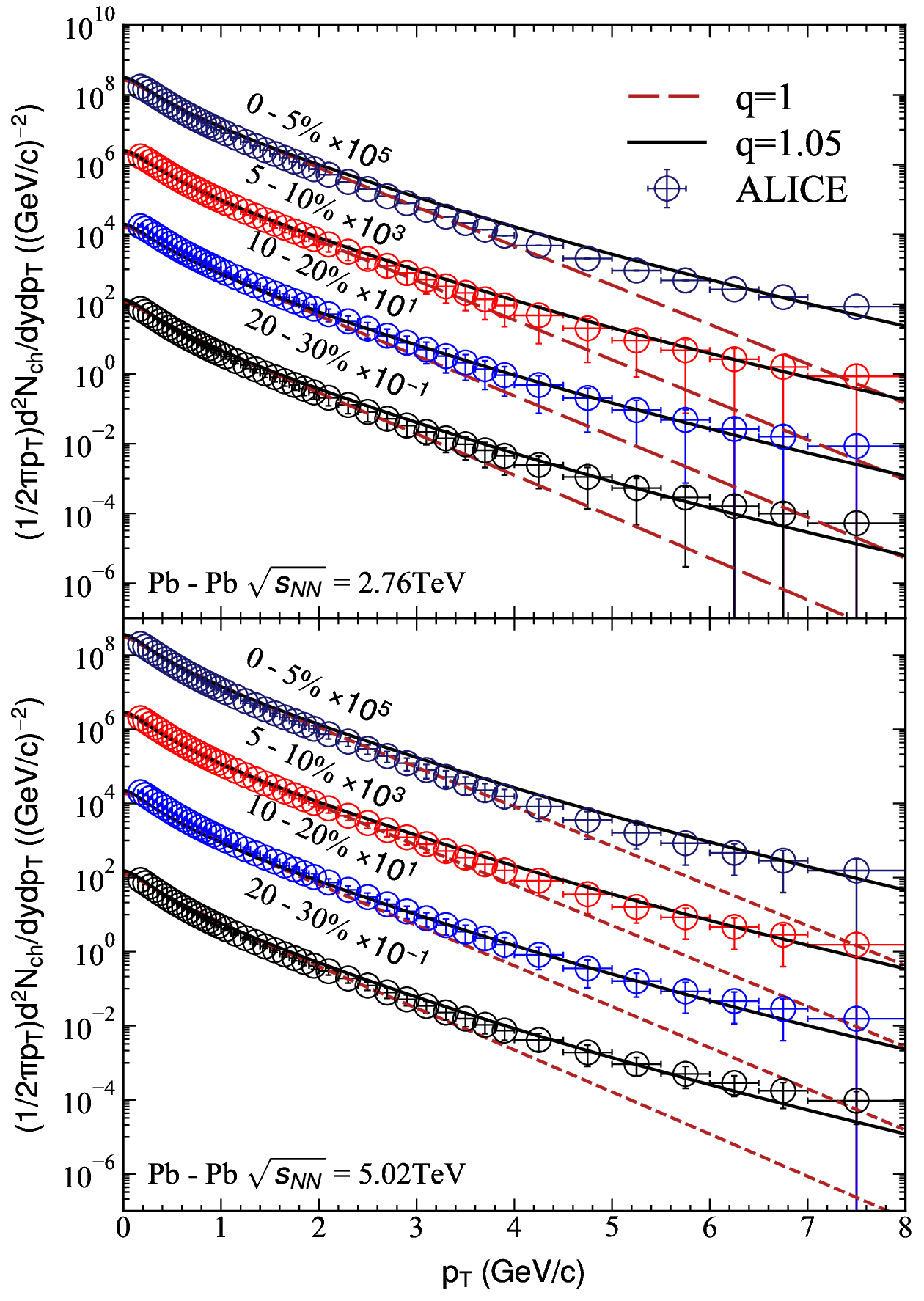}
	\caption{Upper (Lower) panel: $p_{\rm T}$ spectra of charged particles at different centralities in Pb-Pb collisions at $ \sqrt{s_{\rm NN}} $ = 2.76 (5.02) TeV. The solid (dashed) curves are the results from the NEX-CLVisc model with $q=1.05$ ($q=1$). The circle symbols are the experimental data taken from Ref. \cite{lhc_3}.}\label{fig:pt_spectra}
\end{figure}

The $p_{\rm T}$ spectra of charged particles calculated from the NEX-CLVisc model with $q=1.05$ (solid curves)  and $q=1$ (dashed curves) at different centralities in Pb-Pb collisions at $ \sqrt{s_{\rm NN}} $ = 2.76 TeV and 5.02 TeV are shown in Fig. \ref{fig:pt_spectra}. As shown in Ref. \cite{Tsallis_15}, the Tsallis distribution can describe the $p_{\rm T}$  spectra of charged hadrons in AA collisions up to 20 GeV/c. Thus, in principle, by choosing proper parameters, the $q$-hydrodynamic model could be applicable in even higher $p_{\rm T}$ regime. This needs to be examined in the future.  In our present work, we use the “smooth” method to carry out the numerical integration in Eq. (\ref{eq:freeze_out_spectra}) and compute the particle spectra on the freeze-out hypersurface \cite{clvisc}. With this method, the particle spectra up to 8 GeV/c in $N_y\times N_{p_{\rm T}}\times N_\phi=41\times 32\times 48$ tabulated ($y,p_{\rm T},\phi)$ bins are obtained, with $y$ and $\phi$ being the rapidity and azimuthal angle of charged particles. The spectra are then compared to the experimental data, which are presented by empty circles. For better visibility, the spectra are multiplied by the scaling factors indicated in the figure. For both collision energies, the optimized kinetic freeze-out temperature is chosen as $T_f$ = 150 MeV. At a given energy, both the optimum $q$ and $T_f$ are determined by the the fit to the $p_{\rm T}$ spectra of charged particles at the most central collisions. They are then kept fixed for other centrality bins. In the fit, we find that at a fixed $T_f$ the $p_{\rm T}$ spectra become harder with the increase of $q$ . Moreover, at a fixed $q$, the $p_{\rm T}$ spectra  will get steeper with the increase of $T_f$, which is similar to the phenomenon observed in Ref. \cite{pt_flatter}. From the figure, it is found that the extensive result can only describe the data up to 2-3 GeV/c while the non-extensive simulation could reproduce the data up to 6-8 GeV/c. Compared with the parameters in Ref. \cite{eos},  the $q$ value in the non-extensive result is smaller, and the freeze-out temperature is higher. The possible reason is that the lists of decay channels in our work and in Ref. \cite{eos} are different. The consideration with more resonance decays will lead to a smaller $q$ value.

Moreover, our results confirm the conclusion in Ref. \cite{eos} that the value of $q$ is not much affected by the collision energy, as the non-extensive simulation with the same $q$ value gives a nice description of data in central Pb-Pb collisions at $ \sqrt{s_{\rm NN}} $ = 2.76 TeV and 5.02 TeV. However, there is some centrality dependence of the $q$-hydrodynamic description. When going from central to peripheral collisions, the simulation with the same $q$ value is less effective in describing the charged-particle spectrum at high $p_{\rm T}$. This is a reasonable phenomenon, as the degree of non-equilibrium in peripheral collisions is higher than that in central collisions \cite{q_cent_dependence_1, q_cent_dependence_2}.

\begin{figure*}[htpb]
	\centering
	\includegraphics[scale=0.3]{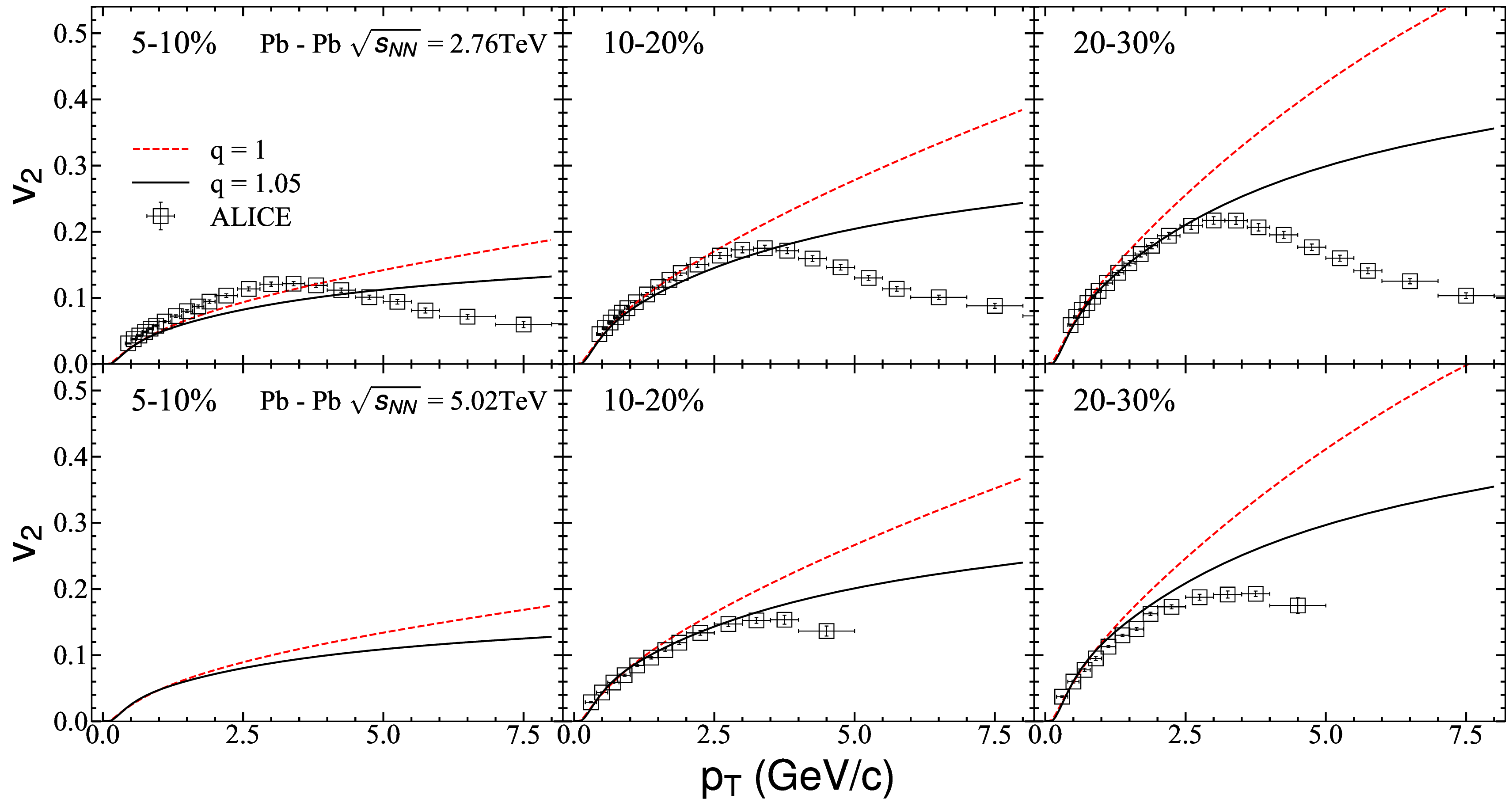}
	\caption{Upper (Lower) panels: the elliptic flow of charged particles at different centralities in Pb-Pb collisions at $ \sqrt{s_{\rm NN}} $ = 2.76 (5.02) TeV. The solid (dashed) curves are the results from the NEX-CLVisc model with $q=1.05$ ($q=1$). The empty squares  are the experimental data taken from Refs. \cite{lhc_6, lhc_7}. The data at the 5-10\% centrality in Pb-Pb collisions at  $\sqrt{s_{\rm NN}} $ = 5.02 TeV is not available so far.}\label{fig:elliptic flow}
\end{figure*}

\subsection{$p_{\rm T}$-differential elliptic flow }
The $p_{\rm T}$-differential elliptic flow $v_2$ of charged particles evaluated from the NEX-CLVisc model with $q=1.05$ (solid curves) and $q=1$ (dashed curves) at different centralities in Pb-Pb collisions at $ \sqrt{s_{\rm NN}} $ = 2.76 TeV and 5.02 TeV are presented in Fig. \ref{fig:elliptic flow}. The experimental data (empty squares) are also shown in the figure. It is observed that the elliptic flow is reduced with the increase of $q$. This suppression of $v_2$ is caused by the modification of the phase-space distribution at the kinetic freeze-out. The effect of the modification is similar as that from the shear viscous corrections, but opposite to that from the bulk viscous corrections \cite{vis_correct}.

Besides the viscous corrections, the elliptic flow is also sensitive to the event-by-event fluctuation. In NEX-CLVisc, the QGP is regarded as an ideal fluid. Moreover, the optical Glauber model is used for the initial condition. It ignores the  fluctuations. Thus the experimental data in the figure are used to guide the eyes, not for quantitative comparison. Even so, one would still observe that with the non-extensive simulation the applicability range of the hydrodynamic model is extended from $p_{\rm T} < 2$ GeV/c to a slightly wider range of $p_{\rm T} \sim$ 3-4 GeV/c in the 10-20\% and 20-30\% centrality bins.  As described in Refs. \cite{recombination_1,recombination_2}, the authors found that parton recombination contributes significantly to the final $v_2$ of all hadrons in the mid-high $3<p_T<8$ GeV/c region. Although the $q$-hydrodynamic model can describe the $p_{\rm T}$ spectra up to 8 GeV/c, it fails to describe the elliptic flow with $p_{\rm T}>$ 4 GeV/c. Thus more efforts are still needed to improve the $q$-hydrodynamic model.

%indicating that the viscous corrections to the hydrodynamic evolution and to the phase-space distribution should be considered \cite{vis_correct}.

\section{Discussions and conclusions}\label{sec:conclusion}

We have developed a non-extensive (3+1)-dimensional hydrodynamic model, NEX-CLVisc, in the framework of  CLVisc.  The non-extensive effect has been consistently considered in the initial conditions set by the optical Glauber model, the equation of state and the hadron kinetic freeze-out procedure. The viscous corrections are turned off to elucidate the pure effect of non-extensive statistics. With this model, we calculate the $\eta$ distribution, the $p_{\rm T}$ spectra and the $p_{\rm T}$-differential $v_2$ of charged particles in Pb-Pb collisions at $\sqrt{s_{\rm NN}}=$ 2.76 TeV and 5.02 TeV, respectively. It is found that the model can reasonably well reproduce the experimental data of the $\eta$ distribution and the charged-particle spectra in a $p_{\rm T}$ range up to 6–8 GeV/c. When compared with the extensive simulation, the $p_{\rm T}$-differential $v_2$ of charged particles is suppressed in the non-extensive simulation, which is similar to that observed in the hydrodynamic model with shear vicious corrections. Moreover, due to the lack of the viscous corrections and the event-by-event fluctuation, the model can only describe the $p_{\rm T}$-differential $v_2$ up to 3-4 GeV/c, which is smaller than the applicable range of the model for the particle $p_{\rm T}$ spectra.

As can be seen from the work in Refs. \cite{non_ext_hydro,eos} and in our paper, despite of the huge success of the extensive hydrodynamics, there is still some possibilities for the non-extensive hydrodynamics in the explanation of the experimental observables in relativistic heavy-ion collisions.

We have also explored the $q$-hydrodynamic description on the $p_{\rm T}$ spectra \cite{pt_spectra_identified_1} and elliptic flow \cite{pt_spectra_identified_2} of identified particles in Pb-Pb collisions at $\sqrt{s_{\rm NN}}=$ 2.76 TeV. It is found that the model can describe the spectra of charged pions and kaons up to 6 GeV/c while overestimates the proton spectra in the whole $p_{\rm T}$ region. Moreover, the model can reproduce the pion and kaon $v_2$ up to $p_{\rm T}\sim$ 2 GeV/c while fails to describe the proton $v_2$. The discrepancy in the proton spectra may indicate that the non-extensivity for protons is less than that for pions and kaons. In Refs. \cite{Tsallis_9, q_cent_dependence_2, meson_baryon_grouping}, the authors have shown that in proton-proton and AA collisions the non-extensivity for baryons is smaller than that for mesons. The decrease of $q$ will lead to a reduction of the proton yield. The present work is our first step to develop the (3+1)-dimensional $q$-hydrodynamic model. We will consider the case that mesons and baryons have different $q$ values in our next work.

It is worth extending the work to turn on the viscous corrections and use the Trento model to set the initial conditions. As the effect of the shear viscous correction is to reduce the $v_2$ while the effect of the bulk viscous correction is to enhance the $v_2$, the shear viscosity over entropy density extracted from data may not go below the KSS bound \cite{KSS_bound} if both the shear and bulk viscous corrections are switched on for the $q$-hydrodynamic model. Moreover, embedding the non-extensive statistics to the fluid with non-vanishing baryon density and examining its validation in nucleus-nucleus collisions at RHIC energies are also on our schedule.

\begin{acknowledgments}
This paper is in part dedicated to commemorating the Nobel laureate T.-D. Lee who made seminal contributions to the study of relativistic heavy ion collisions. We would like to thank T. Hirano for his suggestions during the preparation of the manuscript. Thanks are also given to Chenyan Li, Wenbin Zhao, Guangyou Qin and Shanshan Cao for their valuable discussions. This work is supported by the research fund from the School of Physics and Information Technology at Shaanxi Normal University, by the Scientific Research Foundation for the Returned Overseas Chinese Scholars, State Education Ministry, by Natural Science Basic Research Plan in Shaanxi Province of China (program No. 2023-JC-YB-012) and by the National Natural Science Foundation of China under Grant Nos. 11447024, 11505108 and 12305138.
\end{acknowledgments}

\end{document}